# From G-Factor to A-Factor: Establishing a Psychometric Framework for AI Literacy

Ning Li[1]

lining@sem.tsinghua.edu.cn

Wenming Deng[1]

dwm22@mails.tsinghua.edu.cn

Jiatan Chen[1]

chen-jt24@mails.tsinghua.edu.cn

[1]Tsinghua University

## Abstract

This research addresses the growing need to measure and understand AI literacy in the context of generative AI technologies. Through three sequential studies involving a total of 517 participants, we establish AI literacy as a coherent, measurable construct with significant implications for education, workforce development, and social equity. Study 1 (N=85) revealed a dominant latent factor—termed the "A-factor"—that accounts for 44.16% of variance across diverse AI interaction tasks. Study 2 (N=286) refined the measurement tool by examining four key dimensions of AI literacy: communication effectiveness, creative idea generation, content evaluation, and step-by-step collaboration, resulting in an 18-item assessment battery. Study 3 (N=146) validated this instrument in a controlled laboratory setting, demonstrating its predictive validity for real-world task performance. Results indicate that AI literacy significantly predicts performance on complex, language-based creative tasks but shows domain specificity in its predictive power. Additionally, regression analyses identified several significant predictors of AI literacy, including cognitive abilities (IQ), educational background, prior AI experience, and training history. The multidimensional nature of AI literacy and its distinct factor structure provide evidence that effective human-AI collaboration requires a combination of general and specialized abilities. These findings contribute to theoretical frameworks of human-AI collaboration while offering practical guidance for developing targeted educational interventions to promote equitable access to the benefits of generative AI technologies.

This paper addresses the need to measure AI literacy by presenting three sequential studies. Understanding AI literacy requires situating it within established theoretical frameworks of cognitive measurement and collaborative intelligence. The concept of a general intelligence factor, or g-factor, has fundamentally shaped psychological measurement since its introduction by Spearman (1904), who demonstrated that performance across diverse cognitive tasks correlates positively, suggesting an underlying general ability. This approach to identifying latent constructs that explain performance across multiple domains provides a foundational model for conceptualizing AI literacy (Brand, 1996; Deary, 2000).

Intelligence measurement has evolved significantly, with modern approaches recognizing both general and domain-specific abilities (Kaufman, 2009). As Sternberg et al. (2001) demonstrated, well-designed intelligence assessments have significant predictive validity for real-world outcomes, a principle that guides our approach to AI literacy measurement. The methodological rigor developed in cognitive assessment research over decades offers valuable techniques for constructing reliable and valid measures of complex abilities. Paralleling the g-factor in intelligence research, our study seeks to identify a potential "A-factor" that may represent a general capability for effective AI interaction.

Similarly, research on collective intelligence has established frameworks for measuring group performance and collaborative capabilities (Woolley, 2009). Studies have consistently demonstrated the increasing importance of collaborative approaches in knowledge production (Wuchty et al., 2007) and problem-solving (Gowers & Nielsen, 2009). Hackman (2002) developed comprehensive frameworks for understanding team performance, while Pentland (2008) examined the patterns of interaction that drive successful collaboration. These frameworks are particularly relevant to human-AI collaboration, as they address how individuals coordinate with external systems to enhance performance.

The emerging field of AI literacy builds upon these established constructs while addressing the unique aspects of human-AI interaction. Ng et al. (2021) conceptualize AI literacy as encompassing both technical knowledge of AI systems and the ability to effectively utilize them, distinguishing this from general digital literacy. Long and Magerko (2020) further delineate specific competencies that constitute AI literacy, including the abilities to critically evaluate AI technologies, understand their capabilities and limitations, and effectively communicate with AI systems.

While drawing on these theoretical foundations, our research uniquely focuses on measuring AI literacy as a coherent construct through empirical studies. The first study establishes AI literacy as a measurable construct by revealing a unidimensional latent factor— termed the "A-factor"—representing individuals' capacity to communicate with, evaluate, and guide AI systems effectively. This approach follows established psychometric methodologies (Cattell, 1966) for identifying underlying factors that explain performance across diverse tasks.

The second study refines this measurement tool by examining four key dimensions of AI literacy: communication, creativity, content evaluation, and step-by-step collaboration. Drawing on a more diverse sample, this effort results in an 18-item assessment battery that reliably captures these dimensions while preserving the overall coherence of the A-factor. These dimensions align with the competencies identified in previous AI literacy frameworks (Long & Magerko, 2020) while providing a standardized approach to measurement.

The third study validates the refined instrument in a controlled laboratory setting, demonstrating its predictive value for real-world tasks. This validation approach follows best practices established in intelligence testing (Sternberg et al., 2001) and collective intelligence research (Woolley, 2009), which emphasize the importance of predictive validity for practical applications. Individuals who scored higher on the AI literacy measure performed better on complex, creativity-driven endeavors (e.g., business planning, creative writing). Yet the measure's predictive strength was diminished for tasks requiring specialized domain knowledge (e.g., image design), illuminating its boundary conditions. Moreover, this study identifies several individual difference factors—such as IQ, prior AI experience, and educational background—that significantly predict AI literacy, providing insights into its development and variability.

This research makes significant contributions by systematically defining AI literacy as a distinct construct and developing a reliable tool to measure it. The findings highlight the multidimensional nature of AI literacy and its predictive value in generative AI contexts, offering a foundation for both theoretical advancement and practical application. Additionally, the identification of factors influencing AI literacy provides actionable insights for education and policy aimed at reducing disparities in AI adoption and utilization. By addressing the need for a standardized measure, this paper lays the groundwork for a deeper understanding of human-AI collaboration and the equitable integration of generative AI into diverse domains.

## Methodology & Experimental Design

### Studies Overview

This research consists of three sequential studies, each designed to explore different aspects of AI literacy and its measurement. The multi-study approach follows established practices in psychometric research, which emphasize the importance of establishing construct validity through multiple methods and samples (Campbell & Fiske, 1959). Study 1 aimed to establish the existence of AI literacy as a coherent construct and provided an initial approach to measuring it through factor analysis. Building on Study 1, Study 2 focused on refining the measurement tool by selecting items for a task set designed to assess participants' ability to communicate with AI, provide creative ideas, evaluate AI-generated content, and collaborate step-by-step with AI. Study 3, conducted as a lab experiment, aimed to further validate the effectiveness of the AI literacy measure in real-world tasks and explore factors influencing AI literacy.

**Study 1**

*Sample and Procedure*

This study recruited participants through online questionnaires distributed across various target groups. Participants provided demographic information, including gender, age and educational background, and were informed of the study's incentives: a two-week free GPT-4 account, a personalized AI competency report, and performance-based cash rewards ranging from 30 to 60 RMB. A total of 137 individuals signed up, and the first 120 were invited to participate. Of these, 91 completed the experiment, resulting in a participation rate of 75.8%. After excluding six participants who did not comply with the experimental guidelines, 85 valid samples were retained.

The participants' average age was 22.54 years, with a range from 18 to 34. Among them, 50 were male (58.8%). Most participants (72, or 84.7%) reported prior experience with generative AI. Educational backgrounds varied, including 5 associate degree holders, 61 undergraduates, 16 master's students, and 3 doctoral students. Participants came from a wide range of institutions, from top universities to general institutions. This diversity ensured a representative sample for analysis, which is crucial for establishing the generalizability of psychological constructs (Henrich et al., 2010).

The experiment was divided into three parts, following a structured approach to cognitive assessment. First, participants completed an intelligence test lasting about 10 minutes, which included 18 graphical reasoning questions from the Raven's Progressive Matrices, known for their reliability in measuring fluid intelligence (Raven, 2003). In the second phase, participants were allotted 40 minutes to complete eight simulated generative AI tasks. These tasks evaluated their ability to collaborate with generative AI, including crafting prompts based on specific scenarios, providing feedback on simulated AI responses, and managing complex tasks by clearly defining their roles in collaboration with the AI.

In the final phase, participants were granted access to GPT-4 and tasked with completing content creation tasks, such as designing video content, creating titles, and drafting introductory copy, within 40 minutes. These results were used to validate the effectiveness of the simulated tasks from the second phase, a critical step in establishing the predictive validity of a new measure. To ensure fairness, participants were prohibited from using generative AI tools in the first two stages of the experiment. GPT-4 access was granted only in the final stage for task completion.

*Scoring and Data Analysis*

After collecting the experimental data, rigorous scoring procedures were implemented to ensure accuracy and reliability. Given the superior performance of large language models in text annotation tasks, GPT-4 was primarily used for scoring, supplemented by human evaluations. This dual approach leveraged GPT-4's efficiency in analyzing and evaluating text while incorporating human judgment to ensure precision and nuanced understanding.

In the second phase, where tasks were simpler and responses shorter, GPT-4 handled all scoring. Detailed scoring criteria were developed for each task type to maintain consistency.

GPT-4 independently scored each response three times, yielding Cronbach's alpha values above 0.8, indicating high internal consistency and reliability according to established psychometric standards (Nunnally & Bernstein, 1994).

In the third phase, which involved more complex and lengthy responses, GPT-4 conducted the initial scoring. To enhance the reliability of these results, human evaluations were added, following a multi-rater approach common in creativity assessment. Six crowdsourced workers, each compensated with 100 RMB, assessed the responses based on overall quality, novelty, and usefulness, using a 1 to 10 scale. The scores of crowdsourced workers demonstrated high internal consistency, with Cronbach's alpha values of 0.857, 0.794, and 0.834 for the three dimensions, well above the recommended threshold of 0.7.

During data analysis, it was found that eight participants had not completed all tasks in the second phase. Since these incomplete responses were mostly from later tasks, and task order was fixed, zero imputation was not suitable. Instead, a random forest method was used to impute the missing values, based on participants' completed task scores and those of other participants who completed all tasks.

*Factor Analysis*

To explore whether the eight task scores from the second phase could be explained by underlying latent factors, we conducted an exploratory factor analysis (EFA), a standard approach for uncovering latent variables in psychological research. This analysis aimed to identify potential factors that capture the shared variance among the task scores, thereby assessing the extent to which these tasks measured individuals' overall ability to use generative AI.

Using Principal Component Analysis (PCA) as the extraction method, we identified the factors that explained the maximum variance in the data. The proportion of variance explained by each factor was reported and compared to findings from cognitive intelligence and collective intelligence research, providing a comprehensive context for interpreting the results.

The EFA results revealed that a dominant factor accounted for 44.16% of the variance across the eight tasks, closely aligning with the variance typically explained by the first factor in traditional intelligence tests. A second factor explained only 19.62% of the variance, suggesting a much weaker explanatory role. These findings mirror patterns observed in individual and collective intelligence research, where the first factor generally exhibits strong loadings across all tasks. This dominant factor likely reflects a general underlying ability related to the effective use of generative AI across diverse tasks, similar to the g-factor in intelligence research and the c-factor in collective intelligence.

These results support the hypothesis that a common latent factor underpins participants' generative AI capabilities, validating the experimental tasks as a robust measurement tool. They provide a valuable framework for understanding individual differences in generative AI usage and offer methodological insights for future research. Based on these findings, we extracted the first principal factor and named it the "A-factor," calculating participants' factor scores as a measure of AI literacy.

**Study 2: Refining the Measurement Tool**

*Sample and Procedure*

Study 2 focused on refining our AI literacy measurement through item selection and validation. A total of 286 participants were recruited through online channels, with incentives that included monetary rewards (30-50 RMB), a personalized AI literacy report, and a chance to win one month of GPT-4 access for top performers.

The experiment assessed participants across 40 simulated generative AI tasks designed to measure various facets of AI literacy, including: - Communication effectiveness with AI - Creative idea generation - Content evaluation and feedback capabilities - Problem decomposition for AI collaboration.

The 40 tasks were divided into 5 groups, each containing 16 tasks, with each individual task appearing in two separate groups. This design follows established psychometric approaches to construct validation by ensuring comprehensive coverage while maintaining manageable participant workload. Participants were randomly assigned to one of the 5 groups and were given 60 minutes to complete the tasks.

Task scoring employed GPT-4, following recent evidence that large language models can provide reliable scoring for complex tasks. Each task was scored three times by GPT-4 to ensure reliability, with final Cronbach's alpha exceeding 0.8 for all tasks, indicating high internal consistency.

In terms of missing data handling, 64 participants who failed to complete all tasks were excluded rather than having their data imputed. This decision, while potentially introducing some selection bias, ensured data integrity given the fixed order of tasks and large sample size.

*Analytics and Results*

To evaluate the structure of the 40 tasks, we conducted Exploratory Factor Analysis (EFA) on the five groups separately using Principal Component Analysis (PCA) as the extraction method. The analysis revealed that the first factor explained a significantly larger proportion of variance than subsequent factors in all five task groups, suggesting a dominant, higher-order

factor—referred to as "A-factor"—that captures all aspects of AI literacy. This finding aligns with intelligence research traditions where a general factor (g-factor) explains a significant portion of variance across different cognitive tasks.

Based on the EFA results, we proceeded with item selection following best practices in psychometric test development. We averaged the factor loadings of each task on the first factor across the two groups where it appeared. Items with an average factor loading lower than 0.5 were removed, resulting in a more refined set. For the dimension measuring communication ability with AI, only two items remained, while for the other three dimensions, all items were retained. To further ensure measurement quality, we selected the top five items with the highest factor loadings for the dimensions of creativity, content evaluation, and feedback provision, and the top six items for creative idea generation.

This process led to a final 18-item task set. Importantly, the correlations between items were all below 0.9, ensuring no redundant items, with each item measuring a unique aspect of AI literacy.

**Study 3: Validation of the A-Factor**

*Sample and Procedure*

Study 3 was conducted as a laboratory experiment with 146 participants, providing more controlled conditions than the online studies. Participants were incentivized with 80 RMB in cash, an AI literacy assessment report, and a small gift.

Demographic information collected included age, gender, educational background, prior AI experience, and AI training history. The sample was diverse, with an average age of 24.14 years (SD = 4.76), 30.8% male participants, and varying education levels: 63 participants held undergraduate degrees, 64 had master's degrees, 8 had doctoral degrees, and 11 had associate degrees or lower.

The experiment consisted of three parts, similar to Study 1: 1. Completion of an 18-item Raven test 2. Completion of the 18-item AI literacy task set (75 minutes, without generative AI use) 3. Human-AI collaborative tasks including business plan design, short story writing, and image design (untimed)

One participant completed only the first two phases due to scheduling conflicts, and the image design task was not included in the initial experimental phase. As a result, 145 participants completed the business plan and short story tasks, while 122 participants completed the image design task.

Task order was fully randomized in the second phase, ensuring that incomplete tasks reflected participant ability rather than order effects, a methodological consideration aligned with best practices in psychological assessment. Unfinished tasks were assigned a score of 0.

Scoring procedures mirrored Study 1, with the 18-item task set scored by GPT-4 (three ratings per task, averaged), maintaining high reliability (Cronbach's alpha > 0.8). For the three human-AI collaboration tasks, we employed 22 crowdsourced workers following training on evaluation criteria. The business plan and short story tasks were evaluated based on overall quality (Cronbach's alpha = 0.709 and 0.699, respectively), while image design ratings assessed both artistic quality and topic relevance (Cronbach's alpha = 0.564 and 0.581, respectively).

**Factor Analysis Results**

To explore the underlying structure of AI literacy, we conducted exploratory factor analysis (EFA) using principal component analysis on the 18-item task set. The results revealed that the first factor explained 28.37% of the variance, while the second factor explained only

8.80%. The loadings for the first factor were significantly positive across all items, providing further support for the A-factor hypothesis. This pattern of a dominant first factor is consistent with findings in intelligence research and collective intelligence.

Following the EFA, we performed confirmatory factor analysis (CFA) to validate the factor structure. The CFA model consisted of four first-order factors (representing dimensions of AI literacy) aggregated into a higher-order factor. Using maximum likelihood estimation (MLR), the model fit indices indicated good fit: Comparative Fit Index (CFI) = 0.967 and Root Mean Square Error of Approximation (RMSEA) = 0.028. These values exceed conventional thresholds for good model fit, supporting the structure in which four first-order factors combine into a single higher-order factor representing overall AI literacy.

Based on these analyses, we concluded that the first principal factor extracted in the EFA adequately captures participants' AI literacy, and we calculated participants' factor scores for use in subsequent analyses.

**Regression Analysis**

To validate the effectiveness of A-factor in measuring generative AI usage ability and to explore factors influencing AI literacy, we conducted regression analyses examining both the predictive power of A-factor and its determinants.

First, we examined A-factor's ability to predict performance on three human-AI collaboration tasks. A-factor significantly predicted performance on the short story creation task ($b = 0.717$, $p < 0.001$; with controls: $b = 0.606$, $p < 0.001$). It also positively correlated with business plan performance, though more weakly ($b = 0.301$, $p < 0.05$; with controls: $b = 0.258$, $p < 0.1$). The differential predictive power suggests that AI literacy differences may be more pronounced in more complex tasks.

Interestingly, A-factor did not significantly predict performance on image design tasks, suggesting that visual design with AI may rely more on specialized knowledge where general AI literacy plays a smaller role. This domain specificity in predictive validity parallels findings in intelligence research where general factors may not predict equally well across all specialized domains (Mackintosh, 2011).

Next, we examined factors predicting individual differences in AI literacy. Results showed that AI literacy (A-factor) was significantly associated with: 1. IQ scores ($b = 0.027$, $p < 0.001$; with all controls: $b = 0.020$, $p < 0.001$) 2. Prior generative AI experience ($b = 0.671$, $p < 0.001$; with all controls: $b = 0.340$, $p < 0.1$) 3. AI training history ($b = 0.537$, $p < 0.05$; with all controls: $b = 0.356$, $p < 0.1$) 4. Education level ($b = 0.589$, $p < 0.001$; with all controls: $b = 0.484$, $p < 0.001$)

These findings suggest that both inherent cognitive abilities and acquired knowledge/experience contribute to AI literacy, paralleling research on development of expertise in other domains (Ackerman, 1996).

**Study Results & Findings**

Our research established AI literacy as a coherent, measurable construct through three sequential studies. Study 1 identified a dominant latent factor—termed the "A-factor"—that explained 44.16% of the variance in task performance, substantially exceeding subsequent factors and suggesting a unidimensional structure underlying AI literacy. This emergence of a dominant factor parallels findings in general intelligence research, where the g-factor explains a significant portion of variance across diverse cognitive tasks.

As shown in Table 1, the proportion of variance explained by the first factor in our study (44.16%) was comparable to that observed in collective intelligence tests (43.39%) and higher than in individual intelligence tests (38.77%). This pattern suggests that AI literacy, like collective intelligence, represents a distinct cognitive construct that captures a unique set of abilities related to human-AI interaction.

Correlation analysis revealed significant positive relationships between A-factor scores and task performance measures, including overall performance ($r = 0.23$, $p < 0.05$), novelty ($r = 0.30$, $p < 0.01$), and usefulness ($r = 0.25$, $p < 0.05$). These correlations remained significant even after controlling for IQ, age, and gender in regression analyses, supporting the discriminant validity of AI literacy as distinct from general intelligence.

Study 2 refined the AI literacy measurement through item selection and validation with a larger sample ($N = 286$). Exploratory factor analyses conducted on five task groups consistently revealed a dominant first factor explaining between 31.84% and 47.83% of variance across groups. This pattern of results provides robust evidence for the A-factor as a higher-order construct that encompasses multiple dimensions of AI literacy, similar to hierarchical models of intelligence.

Study 3 validated the refined 18-item AI literacy measure in a controlled laboratory setting ($N = 146$). Exploratory factor analysis of the 18-item task set revealed that the first factor explained 28.37% of the variance, while the second factor explained only 8.80%. Confirmatory factor analysis validated a hierarchical model with four first-order factors aggregated into a higher-order factor, with excellent model fit (CFI = 0.967, RMSEA = 0.028).

The predictive validity of the A-factor was demonstrated through its significant relationship with performance on human-AI collaboration tasks, particularly the short story creation task ($r = 0.46$, $p < 0.001$) and business plan task ($r = 0.20$, $p < 0.05$). The stronger relationship with the short story task suggests that AI literacy's impact may be more pronounced in tasks requiring creative and complex interactions with AI systems.

Interestingly, AI literacy did not significantly predict performance on image design tasks (r = 0.14, p > 0.05 for relevance; r = 0.00, p > 0.05 for artistic quality). This domain specificity highlights the potential boundaries of AI literacy's predictive validity.

**Synthesis of Key Findings**

A consistent finding across all three studies was the emergence of a dominant factor—the A-factor—that explained a substantial portion of variance in AI-related task performance. The stability of this factor across different samples, task sets, and analytical approaches provides strong evidence for AI literacy as a coherent, measurable construct.

While our findings support a higher-order A-factor, they also reveal the multidimensional nature of AI literacy. The four dimensions identified—communication effectiveness, creative idea generation, content evaluation, and problem decomposition—align with theoretical frameworks of human-AI interaction that emphasize the importance of effective communication, creativity, critical evaluation, and strategic thinking in collaborative contexts.

The differential predictive validity of AI literacy across task domains provides insights into its practical utility and limitations. The stronger relationship with creative writing tasks compared to business planning and image design suggests that AI literacy may be particularly important for tasks requiring extensive linguistic interaction and creative collaboration with AI systems.

Our regression analyses identified several significant predictors of AI literacy, including IQ, prior AI experience, AI training, and education level. The significant relationship between IQ and AI literacy suggests that general cognitive ability facilitates adaptation to new technological environments, consistent with theories of fluid intelligence as a foundation for learning new skills. The influence of prior experience and training highlights the role of deliberate practice and structured learning in developing AI literacy, aligning with expertise development theories.

## Discussion

**Theoretical Contributions**

Our research makes several significant theoretical contributions to the understanding of human-AI interaction. First and foremost, we have established AI literacy as a distinct, measurable construct with a coherent factor structure. The emergence of the A-factor across three studies with different samples and methodologies provides robust evidence for the existence of a general ability to effectively interact with generative AI systems. This finding parallels the discovery of the g-factor in intelligence research (Spearman, 1904) and the collective intelligence factor in group performance (Woolley et al., 2010).

The multidimensional yet hierarchical structure of AI literacy revealed in our studies aligns with contemporary theories of intelligence that recognize both general and specific abilities. Similar to how general intelligence encompasses various cognitive abilities while maintaining a coherent structure, AI literacy integrates multiple competencies—communication effectiveness, creative idea generation, content evaluation, and problem decomposition—under a higher-order factor.

Our conceptualization of AI literacy bridges the gap between individual and collaborative intelligence frameworks. Unlike traditional intelligence measures that focus solely on individual cognitive abilities, or collective intelligence measures that examine group dynamics, AI literacy captures the unique cognitive demands of human-AI partnerships. This hybrid nature reflects what scholars have described as "extended cognition," where cognitive processes span the boundary between human and technological systems.

The differential predictive validity of AI literacy across task domains suggests that the construct operates within specific boundary conditions. This domain specificity aligns with theories of situated cognition and transfer of learning, which emphasize the contextual nature of knowledge application.

**Practical Implications**

The identification of AI literacy as a measurable construct has significant implications for education and training. Our finding that AI literacy is influenced by both inherent cognitive abilities (IQ) and acquired knowledge/experience (prior AI experience, training, education) suggests that it can be deliberately developed through structured learning experiences (Bransford et al., 2000). Educational institutions and organizations can design curricula and training programs specifically targeting the four dimensions of AI literacy identified in our research, potentially reducing disparities in AI utilization.

The 18-item assessment tool developed in this research provides a standardized method for evaluating AI literacy, which can be used to assess baseline abilities, track progress, and evaluate the effectiveness of educational interventions. Furthermore, the identification of specific predictors of AI literacy can help educators tailor interventions to different learner profiles, consistent with personalized learning approaches.

As generative AI becomes increasingly integrated into workplace processes, organizations face the challenge of ensuring their workforce can effectively utilize these tools. Our research suggests that AI literacy represents a critical competency for the modern workplace, similar to digital literacy but with distinct characteristics specific to AI interaction.

Organizations can use the AI literacy framework to inform hiring practices, professional development initiatives, and team composition strategies.

The finding that AI literacy significantly predicts performance on complex, creativity-driven tasks has implications for job design and task allocation in AI-augmented workplaces. Organizations might consider matching employees with higher AI literacy to tasks requiring sophisticated human-AI collaboration, while providing additional support or alternative workflows for those with lower AI literacy.

The identification of significant disparities in AI literacy has important implications for policy aimed at ensuring equitable access to the benefits of generative AI. Our findings that education level and prior AI experience significantly predict AI literacy suggest that existing educational and digital divides may translate into AI literacy gaps, potentially exacerbating social and economic inequalities. Policymakers should consider initiatives to promote widespread AI literacy development, particularly among underserved populations.

**Future Research Directions**

While our research establishes AI literacy as a measurable construct and identifies some of its determinants, questions remain about how it develops over time. Future research should employ longitudinal designs to examine the trajectory of AI literacy development across different age groups and in response to various educational interventions.

Our research was conducted within a specific cultural context, and the generalizability of the AI literacy construct across different cultural settings remains an open question. Future research should examine potential cross-cultural variations in AI literacy and its measurement, considering how cultural factors might influence human-AI interaction patterns (Nisbett et al., 2001).

The domain specificity of AI literacy's predictive validity suggests the potential existence of specialized forms of AI literacy for different fields. Future research should explore how AI literacy manifests in specific professional domains such as healthcare, law, education, and creative industries.

While our research focused on individual AI literacy, future studies should examine how AI literacy functions at the group and organizational levels. Research questions might include how teams with varying levels of individual AI literacy collaborate with AI systems, and how organizational structures and cultures influence collective AI literacy development.

**Limitations & Future Research**

While our research provides valuable insights into AI literacy, several methodological limitations should be acknowledged. First, our sampling approach across all three studies may limit the generalizability of our findings. Study 1 and Study 2 relied on online recruitment, which potentially introduced self-selection bias. Participants who volunteered for these studies likely had pre-existing interest in AI technologies, potentially inflating AI literacy scores compared to the general population.

Furthermore, our samples were predominantly drawn from university settings, resulting in a higher average educational level than the general population. This sampling characteristic is particularly relevant given our finding that education level significantly predicts AI literacy. Research based primarily on college student samples may yield findings that do not generalize to broader populations (Henrich et al., 2010). Future research should employ more diverse sampling strategies to ensure representation across different educational backgrounds, age groups, and socioeconomic strata.

The measurement of AI literacy presents several challenges that warrant consideration. Our operationalization of AI literacy through simulated tasks, while ecologically valid, may not capture all dimensions of real-world human-AI interaction. As Messick (1995) argues, construct validity requires evidence that the measurement approach adequately represents the construct domain. While our factor analyses support the internal structure of our measure, additional validation approaches would strengthen confidence in our measurement approach.

Additionally, our reliance on GPT-4 for scoring participant responses introduces potential biases related to the AI system's own limitations and training data. Although we implemented multiple scoring iterations and human verification to mitigate these concerns, the potential for systematic scoring biases remains a limitation of our approach.

The cross-sectional nature of our studies limits causal inferences about the development of AI literacy and its relationship with predictor variables. While our regression analyses identified significant predictors of AI literacy, longitudinal designs would provide more robust evidence regarding the causal relationships between these variables and the development of AI literacy over time.

Our research was conducted within a specific cultural context, potentially limiting the cross-cultural validity of our findings. The conceptualization and measurement of AI literacy may be influenced by cultural factors such as individualism-collectivism, power distance, and uncertainty avoidance, which were not explicitly addressed in our research design.

Building on the limitations identified, future research should focus on expanding and refining the measurement framework for AI literacy. This could include developing and validating domain-specific AI literacy measures for fields such as healthcare, education, creative industries, and business. Such measures would acknowledge the contextual nature of expertise and provide more targeted assessments for specific professional domains.

Additionally, future research should explore alternative measurement approaches beyond task-based assessments, such as self-report measures, observational protocols, and portfolio assessments. A multi-method approach to measuring AI literacy would provide more comprehensive evidence regarding the construct's manifestation across different contexts and assessment formats.

Future research should employ longitudinal designs to examine the developmental trajectories of AI literacy across different age groups and in response to various educational interventions. Such research could address questions regarding the optimal timing for AI literacy development, the stability of individual differences in AI literacy over time, and the effectiveness of different instructional approaches in enhancing AI literacy (Bransford et al., 2000).

Research exploring the ethical and policy implications of AI literacy disparities would be valuable. This could include studies examining how AI literacy relates to algorithmic fairness perceptions, how disparities in AI literacy influence access to economic and educational opportunities, and how policy interventions might address these disparities.

Research on the development and implementation of AI literacy standards and curricula would also be valuable for policy development. Such research could examine questions regarding the appropriate content and sequencing of AI literacy education, the integration of AI literacy into existing educational frameworks, and the assessment of AI literacy at different educational levels.

**Conclusion**

This research represents a significant step forward in understanding and measuring AI literacy in the context of generative AI technologies. Through a series of three sequential studies, we have established AI literacy as a coherent, measurable construct with important implications for education, workforce development, and social equity in an increasingly AI-driven world.

Our research has yielded several important findings that advance our understanding of AI literacy. First, we have demonstrated that AI literacy exists as a coherent construct with a dominant general factor (A-factor) that accounts for a substantial portion of variance in individuals' ability to effectively utilize generative AI tools. Second, our research has revealed that AI literacy is multidimensional, encompassing both general capabilities and domain-specific skills. Third, we have identified significant predictors of AI literacy, including cognitive abilities, educational background, prior technology experience, and personality traits. Fourth, our research has demonstrated the predictive validity of AI literacy for real-world task performance across multiple domains.

The findings from this research have several broader implications for theory, practice, and policy. From a theoretical perspective, our research contributes to emerging frameworks of human-AI collaboration by providing empirical evidence for a distinct form of collaborative intelligence that emerges from human-AI interaction. From a practical perspective, our research provides a foundation for developing educational interventions and training programs aimed at enhancing AI literacy. From a policy perspective, our research highlights the potential for AI literacy disparities to exacerbate existing social inequalities if not addressed through intentional interventions.

As generative AI technologies continue to transform how we work, learn, and create, understanding and developing AI literacy will become increasingly important for individuals, organizations, and societies. Our research provides a foundation for this understanding by establishing AI literacy as a coherent, measurable construct with significant implications for human performance in AI-augmented task environments. By building on this foundation through continued research and practical applications, we can work toward a future in which the benefits of generative AI technologies are widely shared and human-AI collaboration enhances human capabilities rather than diminishing them.

The concept of AI literacy represents a fundamental shift in how we think about human capabilities in the digital age. Rather than viewing AI as a replacement for human intelligence, the AI literacy framework emphasizes the complementary nature of human and artificial intelligence, highlighting the unique capabilities that emerge from their integration. In conclusion, AI literacy represents not just a set of technical skills but a broader capability for effective human-AI collaboration that will be essential for success in the 21st century.

**Figure 1**

*proportion of variance explained in CFA (study1)*

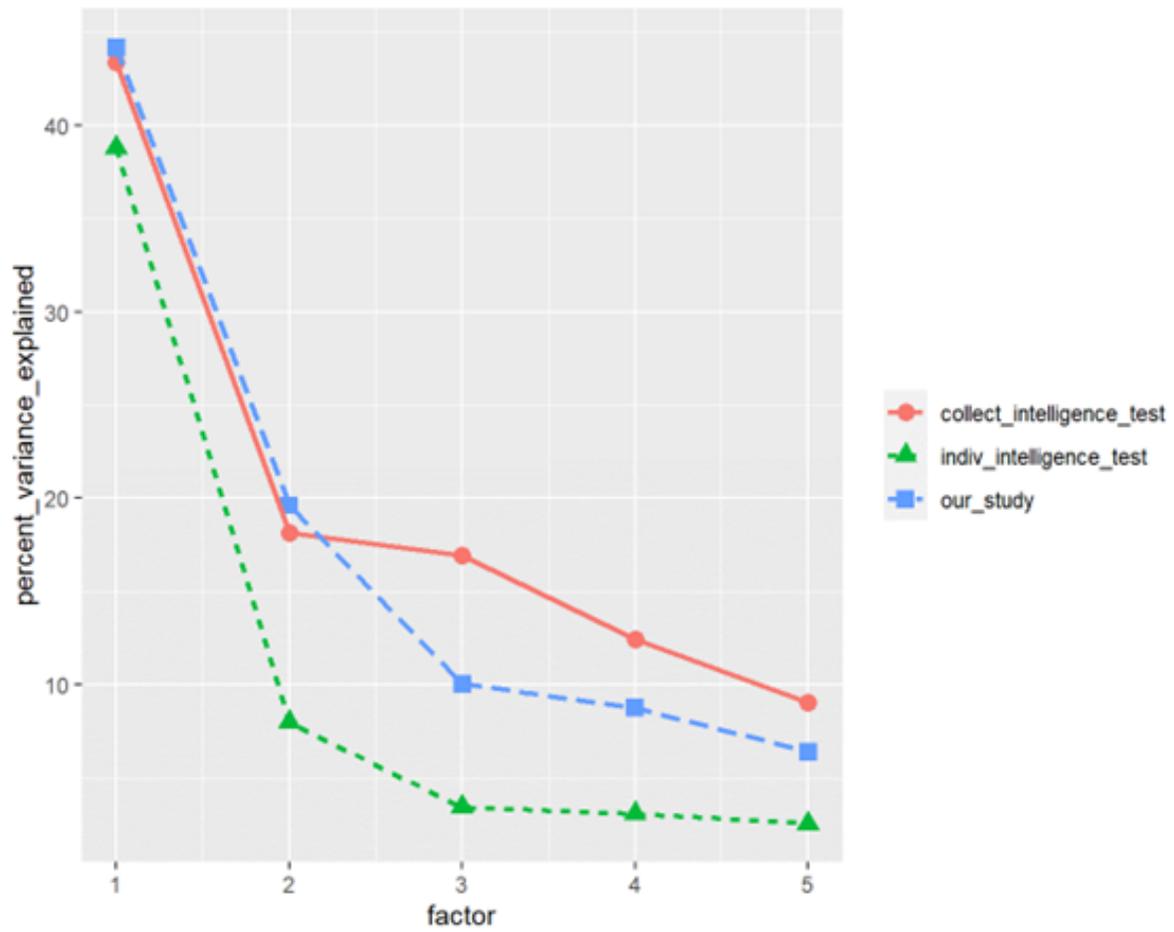

**Table 1**

*proportion of variance explained in CFA (study1)*

|  | Factor 1 | Factor2 | Factor3 | Factor4 | Factor5 |
|---|---|---|---|---|---|
| Collective intelligence test | 43.39 | 18.18 | 16.93 | 12.46 | 9.04 |
| Individual intelligence test | 38.77 | 8.01 | 3.47 | 3.11 | 2.58 |
| Our study | 44.16 | 19.62 | 10.05 | 8.76 | 6.43 |

**Table 2**

*Descriptive Statistics and Correlations of Variables (Study 1)*

| variables | M | S.D. | 1 | 2 | 3 | 4 | 5 | 6 | 7 |
|---|---|---|---|---|---|---|---|---|---|
| 1.AI literacy | 0.00 | 1.000 | | | | | | | |
| 2.overall | 5.46 | 1.363 | 0.23* | | | | | | |
| 3.novelty | 4.90 | 1.230 | 0.30** | 0.89*** | | | | | |
| 4.usefulness | 5.35 | 1.386 | 0.25* | 0.97*** | 0.88*** | | | | |
| 5.age | 22.54 | 3.365 | 0.25* | -0.09 | 0.02 | -0.10 | | | |
| 6.gender | 0.41 | 0.495 | 0.17 | 0.20 | 0.20 | 0.20 | 0.10 | | |
| 7.IQ | 66.76 | 19.128 | 0.19 | 0.23* | 0.22* | 0.25* | 0.19 | 0.26* | |

*Notes:* $^*p < .05$, $^{**}p < .01$, $^{***}p < .001$.

**Table 3**

*Regression models (Study 1)*

| VARIABLES | overall | overall | novelty | novelty | usefulness | usefulness |
|---|---|---|---|---|---|---|
| AI literacy | 0.312* | 0.299* | 0.368** | 0.335* | 0.351* | 0.340* |
|  | -0.15 | -0.15 | -0.13 | -0.13 | -0.15 | -0.15 |
| IQ |  | 0.014† |  | 0.01 |  | 0.015† |
|  |  | -0.01 |  | -0.01 |  | -0.01 |
| age |  | -0.081† |  | -0.034 |  | -0.086† |
|  |  | -0.04 |  | -0.04 |  | -0.04 |
| gender |  | 0.358 |  | 0.315 |  | 0.347 |
|  |  | -0.3 |  | -0.27 |  | -0.3 |
| Constant | 5.463*** | 6.198*** | 4.902*** | 4.907*** | 5.347*** | 6.147*** |
|  | -0.14 | -1.05 | -0.13 | -0.95 | -0.15 | -1.06 |
| Observations | 85 | 85 | 85 | 85 | 85 | 85 |
| R-squared | 0.052 | 0.141 | 0.089 | 0.138 | 0.064 | 0.158 |

Notes: †$p < .10$. * $p < .05$. ** $p < .01$. *** $p < .001$.

**Table 4**

*proportion of variance explained in CFA (study2)*

|           | Factor 1 | Factor2 | Factor3 | Factor4 | Factor5 |
|-----------|----------|---------|---------|---------|---------|
| Task set1 | 45.04    | 9.75    | 8.10    | 7.54    | 6.02    |
| Task set2 | 31.84    | 14.69   | 14.32   | 8.96    | 6.34    |
| Task set3 | 47.83    | 11.60   | 8.30    | 7.30    | 4.67    |
| Task set4 | 41.42    | 11.32   | 8.52    | 6.84    | 5.91    |
| Task set5 | 39.84    | 12.03   | 8.16    | 7.39    | 6.39    |

**Table 5**

*proportion of variance explained in CFA (study3)*

|                    | Factor 1 | Factor2 | Factor3 | Factor4 | Factor5 |
|--------------------|----------|---------|---------|---------|---------|
| 18-item task set   | 28.37    | 8.8     | 6.79    | 6.23    | 5.24    |

**Table 6**

*Descriptive Statistics and Correlations of Variables (Study 3)*

| variables | M | S.D. | 1 | 2 | 3 | 4 | 5 | 6 | 7 | 8 | 9 | 10 | 11 |
|---|---|---|---|---|---|---|---|---|---|---|---|---|---|
| 1.AI literacy | 0.00 | 1.00 | | | | | | | | | | | |
| 2.IQ | 77.74 | 14.43 | 0.40*** | | | | | | | | | | |
| 3.age | 24.14 | 4.76 | -0.03 | -0.05 | | | | | | | | | |
| 4.gender | 1.69 | 0.46 | 0.01 | 0.07 | -0.20* | | | | | | | | |
| 5.everuse | 0.79 | 0.41 | 0.28*** | 0.12 | -0.03 | 0.05 | | | | | | | |
| 6.education | 3.47 | 0.74 | 0.43*** | 0.23** | 0.28*** | -0.02 | 0.22** | | | | | | |
| 7.evertraining | 0.15 | 0.36 | 0.19* | 0.03 | 0.15 | 0.03 | 0.17* | 0.15 | | | | | |
| 8.task1 | 5.17 | 1.46 | 0.20* | 0.13 | 0.08 | 0.03 | 0.05 | 0.13 | 0.06 | | | | |
| 9.task2 | 5.60 | 1.56 | 0.46*** | 0.33*** | 0.00 | 0.03 | 0.31*** | 0.10 | 0.19* | 0.26** | | | |
| 10.task3 pertinent | 6.94 | 1.44 | 0.14 | 0.12 | 0.02 | -0.01 | 0.25** | 0.05 | 0.20* | 0.04 | 0.27** | | |
| 11.task3 artistic | 6.68 | 1.39 | 0.00 | 0.05 | 0.02 | 0.02 | 0.27** | -0.00 | 0.16 | 0.02 | 0.22* | 0.87*** | |

Notes: * $p < .05$. ** $p < .01$. *** $p < .001$.

**Table 7**

*the effectiveness validation of A-factor (Study 3)*

| VARIABLES | task1 | task1 | task2 | task2 | task3 pertinent | task3 pertinent | task3 artistic | task3 artistic |
|---|---|---|---|---|---|---|---|---|
| AI literacy | 0.301* | 0.258† | 0.717*** | 0.606*** | 0.201 | 0.151 | 0.004 | -0.035 |
|  | -0.12 | -0.13 | -0.12 | -0.13 | -0.13 | -0.14 | -0.13 | -0.14 |
| IQ |  | 0.006 |  | 0.019* |  | 0.009 |  | 0.007 |
|  |  | -0.01 |  | -0.01 |  | -0.01 |  | -0.01 |
| age |  | 0.03 |  | -0.001 |  | 0.011 |  | 0.013 |
|  |  | -0.03 |  | -0.03 |  | -0.03 |  | -0.03 |
| gender |  | 0.118 |  | 0.077 |  | -0.041 |  | 0.046 |
|  |  | -0.26 |  | -0.25 |  | -0.29 |  | -0.28 |
| Constant | 5.168*** | 3.790** | 5.590*** | 4.038*** | 6.937*** | 6.013*** | 6.680*** | 5.754*** |
|  | -0.12 | -1.18 | -0.12 | -1.14 | -0.13 | -1.36 | -0.13 | -1.33 |
| Observations | 145 | 145 | 145 | 145 | 122 | 122 | 122 | 122 |
| R-squared | 0.042 | 0.051 | 0.208 | 0.235 | 0.02 | 0.027 | 0 | 0.005 |

Notes: †$p < .10$. * $p < .05$. ** $p < .01$. *** $p < .001$.

**Table 8**

| VARIABLES | AI literacy | AI literacy | AI literacy | AI literacy | AI literacy |
|---|---|---|---|---|---|
| IQ | 0.027*** | | | | 0.020*** |
| | -0.01 | | | | 0 |
| everuse | | 0.671*** | | | 0.340† |
| | | -0.2 | | | -0.18 |
| evertraining | | | 0.537* | | 0.356† |
| | | | -0.23 | | -0.2 |
| education | | | | 0.589*** | 0.484*** |
| | | | | -0.1 | -0.1 |
| age | | | | | -0.029† |
| | | | | | -0.02 |
| gender | | | | | -0.082 |
| | | | | | -0.15 |
| Constant | -2.129*** | -0.529** | -0.081 | -2.040*** | -2.710*** |
| | -0.42 | -0.17 | -0.09 | -0.36 | -0.64 |
| Observations | 146 | 146 | 146 | 146 | 146 |
| R-squared | 0.156 | 0.076 | 0.037 | 0.187 | 0.334 |

*exploring factors influencing AI literacy (Study 3)*

Notes: †$p < .10$. * $p < .05$. ** $p < .01$. *** $p < .001$.